\documentstyle[twoside,twocolumn,psfig,leqno,ico]{article}

\newcommand{\mbi}[1]{\mbox{\boldmath{$#1$}}}

\newcommand{\bgeq}{\begin{equation}}
\newcommand{\edeq}{\end{equation}}
\newcommand{\bgar}{\begin{array}}
\newcommand{\edar}{\end{array}}
\newcommand{\bgea}{\begin{eqnarray}}
\newcommand{\edea}{\end{eqnarray}}
\newcommand{\nnum}{\nonumber}
\newcommand{\nspc}{\\[0.50ex]}

\newcommand{\slsh}{\hspace*{0.05em}--\hspace*{0.05em}}
\begin{document}
\title{Spectral methods in general relativity --- toward the
simulation of 3D\slsh gravitational collapse of neutron stars}
\markright{Spectral methods in general relativity}
\author{S. Bonazzola
\and J. Frieben\thanks{D\'epartement d'Astrophysique Relativiste et de
Cosmologie (UPR 176 du CNRS), Observatoire de Paris, Section de Meudon,
F\slsh 92195 Meudon Cedex, France, {\sl e-mail: frieben@obspm.fr}}\and 
E. Gourgoulhon \and J.A. Marck}
\date{}
\maketitle
\cpr	
%
%
\begin{abstract}
Several applications of spectral methods to problems related to the
relativistic astrophysics of compact objects are presented.
Based on a proper definition of the analytical properties of regular
tensorial functions we have developed a spectral method in a general
sphericallike coordinate system.
The applications include the investigation of spherically symmetric neutron
star collapse as well as the solution of the coupled 2D\slsh Einstein\slsh
Maxwell equations for magnetized, rapidly rotating neutron stars.
In both cases the resulting codes are efficient and give results typically
several orders of magnitude more accurate than equivalent codes based on
finite difference schemes.
We further report the current status of a 3D\slsh code aiming at the
simulation of non\slsh axisymmetric neutron star collapse where we have
chosen a tensor based numerical scheme. 
\end{abstract}
%
%
%
%
\section{Introduction}
Compact objects in astrophysics such as neutron stars and black holes are
subjected to the strong field regime of gravitation and have hence to be
treated within the framework of general relativity. The growing interest in
the numerical solution of the Einstein equations for astrophysically relevant
systems has given rise to a new branch of computational physics --- {\em
numerical relativity\/} \cite{SM79,HB89,IN92}.
This development is due to the increasingly powerful computational resources
which make these problems accessible to a numerical investigation.
It is further stimulated by the prospects of gravitational wave astronomy
which will turn into an observational science toward the end of this decade
thanks to gravitational wave observatories like LIGO, VIRGO and GEO600 that
are now under construction \cite{AB92,BR90,HD94}.\nspc
We use the (3+1)\slsh formalism of general relativity \cite{SY78-2} which
consists in foliating spacetime into a sequence of spacelike hypersurfaces
which represent curved three\slsh space at a fixed coordinate time $t$.
The fabric of spacetime is then determined by the three\slsh metric $h_{ij}$
and four additional quantities, the {\em lapse function} $N$ and the {\em
shift vector} $N^i$ which fix the propagation of the spacelike hypersurfaces
in time and the change of the spatial coordinate system between adjacent
hypersurfaces.
This Hamilton type approach to general relativity results in a temporal first
order evolution scheme for the dynamical variables which is completed by some
constraint equations which ensure the consistency of gravitational and matter
fields. Furthermore $N$ and $N^i$ have to be determined by the choice of
appropiate gauge conditions which typically lead to elliptic equations that
have to be solved at each time step.
For stationary configurations all time derivatives vanish and one obtains a
system of coupled elliptic equations for the gravitational fields.
The efficient solution of elliptic equations is hence of central interest for
us.\nspc
Let us consider a covariant Poisson equation $N^{|i}{}_{|i}\!=\!S$ in a
conformally flat axisymmetric space where the line element reads
\bgeq
dl^2\!=\!A^4(r,\theta)\,(dr^2\!+\!r^2\,d\theta^2\!+\!r^2\sin^2\!\theta\,
d\phi^2).\edeq
The former equa\-tion can be rewritten to yield a Poisson\slsh like equation
for $N$ where we have isolated the flat space Laplacian $\Delta_f$ and
contributed the curvature terms to the source. Here $\alpha$ denotes $\ln A$.
\bgeq
\quad\;\Delta_f N=\tilde{S}\;\;\mbox{with}\;\;\tilde{S}=A^4 S-2\,
(\partial_r\alpha\partial_r N+\frac{1}{r^2}\,\partial_\theta\alpha
\partial_\theta N).
\edeq
This equation has to be solved by iteration. The solution of $\Delta_f N\!=\!
\tilde{S}$ at each iteration has hence to be accomplished sufficiently fast
in order to keep the total computation cost at a reasonable level.\nspc
After outlining the basic features of our spectral method \cite{BM86,BM90},
we will proceed in a first step to the investigation of black hole formation
due to spherically symmetric neutron star collapse which has proved
the high aptitude of spectral methods in this field \cite{GO91,GH93,GHG95}.
The second part is devoted to the study of axisymmetric stationary rotating
bodies which has been applied to model rapidly rotating neutron stars
\cite{BGSM93,SBGH94}.
This work has been extended recently to include strong magnetic fields for
the first time into neutron star models \cite{BBGN95}.
Special emphasis in all cases has been put on the extensive use of external
and intrinsic tests \cite{BO73,BG94,GB94} of the self\slsh consistency and
the attained accuracy of the numerical results.
The resulting neutron star models provide us with the required initial value
models for the investigation of 3D\slsh gravitational collapse of neutron
stars which will reveal the whole range of gravitational wave emission
associated with this phenomenon.
We give an overview about the inset of spectral methods in this project which
is currently in work. Here a new method for the efficient inversion of a
generalized 3D\slsh vector Poisson equation is a first major result.
\section{Spectral methods in general relativity}
\subsection{Coordinates and regularity conditions}\label{SEC:CR}
The spacelike hypersurfaces stemming from the former choice of the
(3+1)\slsh formalism are conceived to describe some asymptotically flat space,
containing a compact, most\-ly starlike object. The natural choice is thus a
sphe\-ri\-callike coordinate system $(r,\theta,\phi)$.\nspc
The pseudosingularities which appear in this case can be overcome by a proper
definition of regularity conditions of tensorial quantities.
A consequent application of parity rules derived from these conditions allows
further to optimize code efficiency and precision.\nspc
We consider the related Cartesian type coordinate system $(x,y,z)\!=\!
(r\sin\theta\cos\phi,r\sin\theta\sin\phi,r\cos\theta)$.
We define a tensorial quantity $T_{i_1\ldots i_N}$ in spherical coordinates
$(r,\theta,\phi)$ to be {\em regular}, if its components $f_{i_1\ldots i_N}$
with respect to Cartesian coordinates $(x,y,z)$ are regular in the sense that
they can be expanded into a polynomial sum of the type
\bgeq\label{EQ:REC} f_{i_1\ldots i_N}(x,y,z)=\sum_{i,j,k=0}^{N}a_{i_1\ldots
i_N ijk}\,x^i y^{\,j} z^k\edeq
which can be written in terms of $(r,\theta,\phi)$ as
\bgea\label{EQ:REG}\tilde{f}_{i_1\ldots i_N}(r,\theta,\phi) & = &
\sum_{i,j,k=0}^{N}a_{i_1\ldots i_N ijk}\,r^{i+j+k}\\
\nnum & &\times\sin^{i+j}\!\theta\,\cos^k\!\theta\,\cos^i\!\phi\,
\sin^j\!\phi.\edea
Having specified the tensor components $f_{i_1\ldots i_N}$ with respect to
the Cartesian frame we derive the components related to the local orthonormal
frame of spherical coordinates by a {{\em non\slsh singular\/} coordinate
transformation.\nspc
In order to infer the analytical properties of a scalar function it is useful
to rearrange the sum in (\ref{EQ:REG}).
We first collect all the terms referring to $\cos m\phi$ and $\sin m\phi$
respectively. We write
\bgeq\label{EQ:SX} \quad\tilde{f}(r,\theta,\phi)=\sum_{m=0}^M\;
(a_m(r,\theta)\,\cos m\phi+b_m(r,\theta)\,\sin m\phi)\edeq
where $a_m(r,\theta)$ and $b_m(r,\theta)$ behave identically in the further
procedure. We opt for $\cos l\theta$ and $\sin l\theta$ as basis functions
in $\theta$ which allows the application of FFT\slsh techniques for this
transformation.
An immediate conclusion from (\ref{EQ:REG}) and the case $i\!+\!j$ even is
that the coefficients $a_{2m}(r,\theta)$ and $b_{2m}(r,\theta)$
have to be expanded in terms of $\cos l\theta$ while from the odd case that
the expansion of $a_{2m+1}(r,\theta)$ and $b_{2m+1}(r,\theta)$ has to be done
on the set $\sin l\theta$. We therefore specify
\bgeq a_{2m}(r,\theta)=\sum_{l=0}^{L}\,\tilde{a}_{l,2m}(r)\,\cos l\theta,\edeq
\bgeq a_{2m+1}(r,\theta)=\sum_{l=0}^{L}\,\tilde{a}_{l,2m+1}(r)\,\sin l\theta.
\edeq
Note that due to the well defined parity of $a_{2m}$ and $a_{2m+1}$ these 
coefficients --- a priori only defined for $0\!\leq\!\theta\!\leq\!\pi$ ---
can be continued analytically to periodic functions of $\theta$ on the
interval $[-\pi,\pi]$.\nspc
In the same manner as before we find from (\ref{EQ:REG}) that the polynomials
$\tilde{a}_{l}(r)$ are symmetric with respect to the inversion $r\rightarrow
-r$ for $l$ even and antisymmetric for $l$ odd. As basis function set in $r$
we decide for Chebyshev polynomials due to their superior properties in
finite approximation schemes of non\slsh periodic functions and the
availability of fast Chebyshev transforms. Simplifications of the expansion
scheme in the presence of additional symmetries are precised in the following
paragraphs.\nspc
Any regular function admits an expansion of this kind, but regularity
furthermore implies additional constraints on the different coefficients.
Having set up regular initial data according to (\ref{EQ:REC}), regularity
of the involved quantities is maintained during a calculation by the
application of regular operators --- we here ignore the influence of numerical
effects due to aliasing or roundoff errors.\nspc
Let us consider the covariant derivative of a vector in sphe\-ri\-cal
coordinates. For the scalar potential
$U(r,\theta,\phi)\!=\!r\sin\theta\cos\phi$ and
\bgeq\label{EQ:XP} U_{|i}=(\partial_r\,U,\frac{1}{r}\,\partial_\theta\,U,
\frac{1}{r\sin\theta}\,\partial_\phi\,U)\edeq
we have
\bgeq\label{EQ:XU} (U_r,U_\theta,U_\phi)=(\sin\theta\cos\phi, \cos\theta
\cos\phi, -\sin\phi).\edeq
The covariant derivative $U_{\theta|\phi}$ which reads
\bgeq\label{EQ:XV} U_{\theta|\phi}=\frac{1}{r\sin\theta}\,\partial_\phi\,
U_\theta-\frac{1}{r\tan\theta}\,U_\phi\edeq
can be rewritten to yield
\bgeq\label{EQ:XT} U_{\theta|\phi}=\frac{1}{r\sin\theta}\,(\partial_\phi\,
U_\theta-\cos\theta\,U_\phi).\edeq
A numerical evaluation according to (\ref{EQ:XP}) and (\ref{EQ:XT}) reveals a
perfectly regular behaviour.
\subsection{Supersymmetric case}\label{SUBSEC:SS}
Additional spatial symmetries of physical systems involve continuous symmetry
operations like rotation about a distinct axis with an associated
Killing vector field or discrete transformations like inversion at the
equatorial plane $z\!=\!0$, leading to distinct parity properties of the
different tensor components.\nspc
We define {\em supersymmetry\/} \cite{MA95} by the following behaviour of a
scalar function:
$f$ is invariant with respect to inversion at the $z-$axis, hence 
$f(-x,-y,z)\!=\!f(x,y,z)$, while $f$ is [anti\slsh ] symmetric with respect
to reflection at the equatorial plane $z\!=\!0$, hence $f(x,y,-z)\!=\!\pm
f(x,y,z)$. Consequently $f$ can be expanded into a sum of the type
\bgeq\label{EQ:SN}\hspace*{1.85em}f_+(x,y,z)=\sum_{l=0}^L\sum_{k=0}^{2l}
\sum_{m=0}^M c_{klm} x^{2l-k}y^k z^{2m}\hspace*{0.5em}\mbox{even case}\edeq
\bgeq\hspace*{1.45em}f_-(x,y,z)=\sum_{l=0}^L\sum_{k=0}^{2l}\sum_{m=0}^M
c_{klm} x^{2l-k}y^k z^{2m+1}\hspace*{0.25em}\mbox{odd case}\edeq
which defines subsets of the general scalar functions introduced in
Sec.~\ref{SEC:CR}.
A write\slsh up in terms of $(r,\theta,\phi)$ gives
\bgea\label{EQ:SM}\tilde{f}_+(r,\theta,\phi)
=\sum_{l=0}^L\sum_{k=0}^{2l}\sum_{m=0}^M c_{klm}\,r^{2(l+m)}\\
\nnum\times\sin^{2l}\!\theta\,\cos^{2m}\!\theta\,\cos^{2l-k}\!\phi\,
\sin^k\!\phi,\\
\tilde{f}_-(r,\theta,\phi)
=\sum_{l=0}^L\sum_{k=0}^{2l}\sum_{m=0}^M c_{klm}\,r^{2(l+m)+1}\\
\nnum\times\sin^{2l}\!\theta\,\cos^{2m+1}\!\theta\,\cos^{2l-k}\!\phi\,
\sin^k\!\phi
\edea
which can be modified, replacing $\sin^2\!\theta$ by $(1\!-\!\cos^2\!\theta)$.
\bgea\label{EQ:SO}\tilde{f}_+(r,\theta,\phi)
=\sum_{l=0}^L\sum_{k=0}^{2l}\sum_{m=0}^M c_{klm}\,r^{2(l+m)}\\
\nnum\times(1\!-\!\cos^2\!\theta)^l\cos^{2m}\!\theta\,\cos^{2l-k}\!\phi\,
\sin^k\!\phi,\\
\tilde{f}_-(r,\theta,\phi)=\sum_{l=0}^L\sum_{k=0}^{2l}\sum_{m=0}^M c_{klm}\,
r^{2(l+m)+1}\\
\nnum\times(1\!-\!\cos^2\!\theta)^l\cos^{2m+1}\!\theta\,\cos^{2l-k}\!\phi\,
\sin^k\!\phi.\edea
According to Sec.~\ref{SEC:CR} we conclude for the decomposition of a
supersymmetric function:
\begin{enumerate}
\item $f$ is $\pi-$periodic in the angular variable $\phi$,
\item $a_{2m}(r,\theta)$ has to be expanded into a sum of $\cos l\theta$
where $l$ is even in the symmetric case and odd in the antisymmetric one,
\item $\tilde{a}_l(r)$ is an even polynomial in $r$ for $f$ symmetric and
an odd polynomial for $f$ antisymmetric.
\end{enumerate}
Physical problems which imply the use of supersymmetric functions allow to 
restrict the computational domain to $[0,\pi]$ in $\phi$ and to
$[0,\pi/2]$ in $\theta$ which leads to an overall reduction of the effective
grid size by a factor four.\nspc
For the components of a vector field associated with the case of even 
supersymmetry we conclude:
$U_r$ can be expanded into a sum of $\cos 2l\,\theta$, $U_\theta$ into a sum
of $\sin 2l\,\theta$ and $U_\phi$ into a sum of $\sin(2l\!+\!1)\,\theta$
which means that $U_\theta$ undergoes a change of sign by reflection at the
equatorial plane while $U_r$ and $U_\phi$ remain unchanged. The expansion of
the radial part is done in terms of odd powers of $r$ for all components.
\subsection{Axisymmetric case}\label{SUBSEC:AX}
Axisymmetry restricts the set of scalar functions under consideration to
functions according to Sec.~\ref{SEC:CR} where the only remaining term in
(\ref{EQ:SX}) is $a_0(r,\theta)$. Thus $f$ can be written as
\bgeq\label{EQ:AX} f(r,\theta)=\sum_{l=0}^{L}\,\tilde{a}_{l,0}(r)\,\cos
l\theta\edeq
where $\tilde{a}_{l,0}(r)$ is an even function in $r$ for $l$ even and an odd
function for $l$ odd.\nspc
For the components $U_r$ and $U_\theta$ of a vector field compatible with the
assumption of axisymmetry we conclude:
$U_r$ has to be expanded in terms of $\cos l\theta$ where the radial part is
even for $l$ odd and odd for $l$ even. $U_\theta$ has to be expanded in terms
of $\sin l\theta$ where we also have a parity change in $r$.\nspc
The properties of the lacking component $U_\phi$ which we need to 
handle $N_\phi$ in the models of rotating neutron stars can be derived
from the Killing equation linked to axisymmetry.
A short examination reveals that $U_\phi$ has to be expanded in terms
of $\sin l\theta$ with a parity change in $r$ --- it behaves identically
as $U_\theta$.
\subsection{Spherically symmetric case}\label{SUBSEC:SP}
For sake of completeness we add the spherically symmetric case.
As a further restriction of the axisymmetric case we keep from
(\ref{EQ:AX}) only $\tilde{a}_{0,0}(r)$. $f(r)$ therefore reads
\bgeq\label{EQ:SP}f(r)=\sum_{k=0}^K\,a_{2k}\,r^{2k}\edeq
which is an ordinary even polynomial in $r$ while $U_r$ is represented by an
odd polynomial in $r$.
\section{Spherically symmetric neutron star collapse}\label{SEC:NS}
\subsection{Basic equations}\label{SUBSEC:AXE}
The investigation of neutron star equilibrium configurations in spherical
symmetry was the first problem in the fully general relativistic regime being 
solved by us by means of the (3+1)\slsh formalism of general relativity and a
spectral method \cite{GO91,BM86,BM90}.
A favourable choice of the line element $ds^2\!=\!g_{\alpha\beta}\,dx^\alpha
dx^\beta$ in the case of spherical symmetry is given
by RGPS ({\em Radial Gauge\slsh Polar Slicing\/}) coordinates \cite{BP83}
and reads
\bgeq ds^2=-N^2 dt^2+A^2 dr^2+r^2\,(d\theta^2\!+\sin^2\!\theta\,d\phi^2).\edeq
We stress the particular nature of this problem where the field variable
$A$ is not really a dynamical quantity. In fact it is uniquely determined at
any moment as well as the lapse function $N$ by the matter fields which have
to be evolved by means of the hydrodynamical equations. Since the solution
outside the star is known in advance to coincide with the {\em static\/} 
Schwarzschild solution of a point mass of the same size according to the
{\em Birkhoff theorem\/}, we benefit from a double simplification.
First the whole time evolution is yet determined by propagating the
hydrodynamical variables and further the calculation can be restricted to the
stellar interior.
The interior solution for the gravitational field has then to be matched to
the analytical exterior one. We further note that due to the static character
of the exterior solution no gravitational waves --- which otherwise would be
an observable of most importance --- are emitted.\nspc
Concerning the hydrodynamical part we employ a set of particular variables
which lead to equations ressembling very closely their Newtonian counterparts
including a general relativistic generalization of the classical Euler
equation.
We note the privileged role of the Eulerian or local rest observer
${\cal O}_0$ in this formulation. The hydrodynamical equations read
\bgeq\partial_t{\cal E}+\frac{1}{r^2}\,\partial_r\left(r^2({\cal E}\!+\!p)
V^r\right)=0,\edeq
\vspace*{-4.0ex}
\bgea\quad\partial_t U + V^r \partial_r U & = &-\frac{1}{{\cal E}\!+\!p}
\left(\frac{N}{A}\,\partial_r p + U\partial_t p\right)\\\quad&&-\frac{AN}
{{\mit\Gamma\/}^2}\left(\frac{m}{r^2}+4\pi rp\right),\nnum\edea
\bgeq\partial_t D + \frac{1}{r^2}\,\partial_r (r^2 D V^r)=0,\edeq
\bgeq\partial_t s_B + V^r \partial_r s_B=0\edeq
where we have introduced the energy density ${\cal E}$ and the fluid velocity
$U$ measured by ${\cal O}_0$, the coordinate baryon density $D$ and the
entropy per baryon $s_B$ while $V^r$ denotes the fluid coordinate velocity.
The Lorentz factor ${\mit\Gamma\/}$ is defined as ${\mit\Gamma\/}\!=
\!(1-U^2)^{-\frac{1}{2}}$. In addition one has to solve
\bgeq\partial_r m(r,t) = 4\pi r^{2}\,{\cal E}(r,t),\edeq
\bgeq\qquad\partial_r \Phi(r,t) = A^2 \left(\frac{m(r,t)}{r^{2}}+4\pi r\,
[\,p+({\cal E}\!+\!p)\,U^2\,]\right)\edeq
where $A(r,t)$ is related to $m(r,t)$ through
\bgeq A(r,t) = \left(1-\frac{2m(r,t)}{r}\right)^{-\frac{1}{2}}.\edeq
The neutron star matter is modeled as a perfect fluid, adopting a realistic
dense matter equation of state.
\begin{figure}[t]
\unitlength 1mm
\psfig{figure=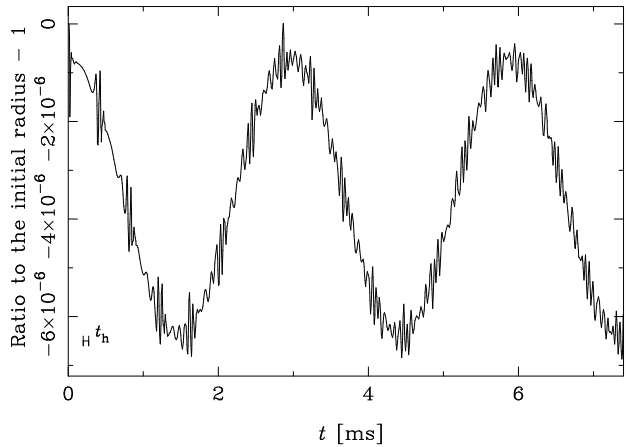,angle=270,width=95mm}
\caption[]{\label{FIG:FMB} Relative variation of the stellar radius with
elapsing time for a stable equlibrium model with $n_{\rm{B}}\!<\!
n_{\rm{B}}^{\rm{crit}}$ and $|\Delta M/M_{\rm{max}}|=-5\!\times\!10^{-5}$.
The hydrodynamical timescale is indicated on the lower left.}
\end{figure}
\subsection{Numerical method}
The initial value model for the dynamical calculations is provided by solving
the Tolman\slsh Oppenheimer\slsh Volkoff equations describing a spherically
symmetric static star where each model is determined by the central value of
the pseudoenthalpy $H_{\rm{c}}$. A first solution is obtained by integration
of this system of ordinary differential equations while an overall numerical
accuracy of the order of $10^{-14}$, adapted to the subsequent use of a
spectral method, is achieved by iteration of the approximate solution.\nspc
The computational domain is identical with the stellar interior during the
whole calculation, thanks to a comoving grid whose outer boundary coincides
with the star surface. This maintains a constant spatial resolution during
the collapse and a fine sampling of the steepening gradients near the star
surface thanks to the accumulation of the Gau\ss\slsh Lobatto points in this
region. It furthermore minimizes the advective terms, hence improving the
numerical accuracy. All quantities are expanded in terms of Chebyshev
polynomials in $r$ mapping the interval $[0,R_*(t)]$ onto $[0,1]$ taking
into account their analytical properties according to Sec.~\ref{SUBSEC:SP}.
The 2$^{\rm nd}$ order semi\slsh implicit time integration ensures the
stability of the code which allows us to perform simulations of a duration
of many dynamical timescales. This ability is very important when studying
the effects of perturbations on equilibrium configurations. This task is
also favoured by the fact that we integrate the original system of equations
without any artificial viscosity to stabilize the code while in addition the
intrinsic viscosity of spectral methods is negligible.
The ingoing characteristic of the hydrodynamical system at $r\!=\!R_*$ gives
rise to one boundary condition which is chosen to fix the baryon density at
the star boundary. It is imposed by means of a $\tau$\slsh Lanczos procedure
\cite{LA56} on the system as a whole which is the well posed mathematical
procedure.
\begin{figure}
\unitlength 1mm
\psfig{figure=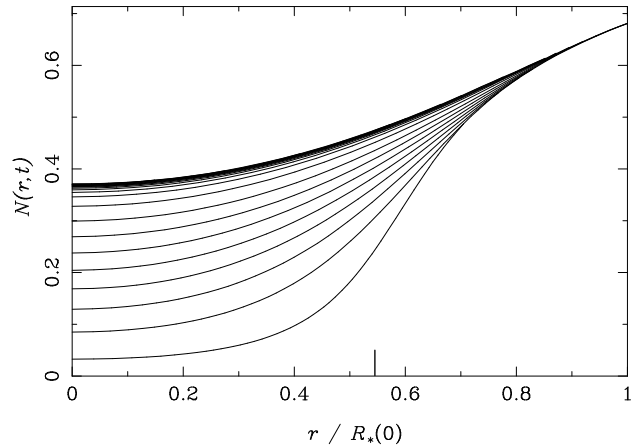,angle=270,width=95mm}
\caption[]{Profiles of the lapse function $N$ for $t$ ranging from 0 to 7.296
ms. The location of the Schwarzschild radius $R_{\mbox{s}}$ is indicated by a
vertical bar. Direction of increasing time is downward.}\label{FIG:NCN}
\end{figure}
\subsection{Results}\label{SUBSEC:NSR}
Among the various tests we have imposed on our code one describes the collapse
of a homogeneous dust sphere whose solution was given by Oppenheimer and Snyder
\cite{OS39}.
From the beginning of the collapse until the moment where the whole
configuration is highly relativistic and practically frozen we observed the
different variables to reproduce the analytical values within an error of
better than $10^{-5}$ \cite{GO91}, while the errors related to previous
studies based on finite difference methods are of the order of $10^{-2}$
\cite{SC89}.\nspc
Neutron star models near the maximum mass configuration --- $M_{\rm{max}}\!=
\!1.924 M_\odot$ and $R\!=\!10.678$ km for the employed EOS --- are interesting
with respect to their stability against radial perturbations. Configurations
with a central baryon density approaching the critical one exihibit an
oscillatory behaviour which is dominated by the {\em fundamental\/} mode of
oscillation of increasing period length.
It represents a uniform growing and shrinking of the entire star which is
modulated by less important harmonics of higher order.
Fig.~\ref{FIG:FMB} shows this temporal variation for a stable configuration.
We stress that these oscillations are entirely driven by roundoff and
discretization errors of a total order of $10^{-10}$ --- no external force has
been applied to trigger this variability.
Increasing the central baryon density beyond the critical value one enters the
branch of unstable configurations. This time the fundamental mode starts a
contraction of the star which results in an unlimited collapse. Though the
actual coordinate choice is not capable to properly capture the formation of
an apparent horizon which would clearly reveal the formation of a black hole
the evolution of metric potentials and matter variables gives a distinct
indication of this event.
Take for instance Fig.~\ref{FIG:NCN} which shows the time development of the
lapse function $N$, measuring the elapsed proper time of the local Eulerian
observer ${\cal O}_0$. Inside the Schwarzschild radius it tends toward zero
for increasing coordinate time $t$. This behaviour --- called the {\em `lapse
of the lapse'\/} --- is related to the singularity avoidance property of the
chosen coordinate gauge which keeps the spatial hypersurfaces from propagating
into a forming singularity. The dynamical timescale of the collapse is the
time elapsed from the beginning of the collapse until the moment where the
evolution appears to be frozen to a distant observer and has the value
$t_{\rm{}}\!=\!7.4$ ms for the considered configuration.
Fig.~\ref{FIG:BCN} illustrates the relative error committed on the total
baryon number which is a conserved global quantity. In the early phase it is
conserved with a relative accuracy of $2\!\times\!10^{-8}$ and with
$4\!\times\!10^{-6}$ during the violent stages of the collapse. The deviation
increases up to $5\!\times\!10^{-5}$ in the final phase where sharp gradients
form near the horizon. The proper working of the code in the perturbative
regime has been recently confirmed by direct comparison with a linear adiabatic
code \cite{GHG95}. The calculations have shown a very good agreement of the
frequencies of the fundamental modes for the two opposite approaches.
\begin{figure}
\unitlength 1mm
\psfig{figure=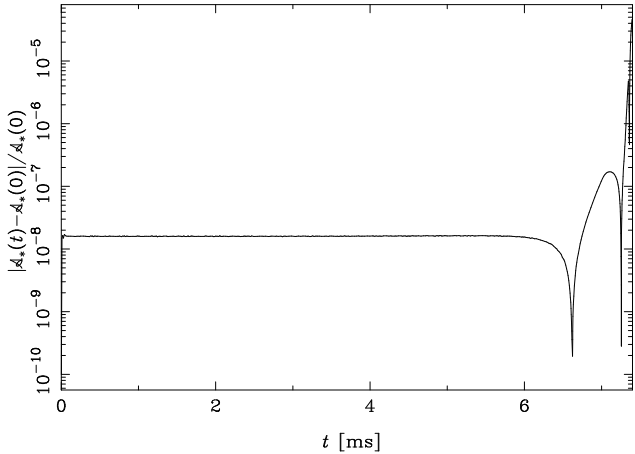,angle=270,width=95mm}
\caption[]{Relative variation of the total baryon number during the collapse.}
\label{FIG:BCN}
\end{figure}
The original version of our code has been extended later on to include the
processes of neutrino production and transport during neutron star collapse
in order to compute the observable neutrino emission for a distant observer
\cite{GH93}.
A multidomain extension of this code is in use in order to simulate
type II supernovae in spherical symmetry. It is particularly well suited to
properly capture the high contrast up to $\simeq 10^6$ of the mean densities
in the individual subdomains which appears during the collapse. Four to five
zones are typically needed to cover the dense central core forming the new
born neutron star and the outer layers of much lower density with the desired
resolution. In contrast with the original code boundary conditions are imposed
in this case by means of a modified $\tau$\slsh Lanczos scheme applied in
{\em coefficient} space. 
\section{Axisymmetric rotating relati\-vi\-stic bodies}
\subsection{Basic equations}\label{SUBSEC:BEMS}
In order to study rapidly rotating neutron stars we have made the assumptions
of spacetime being {\em axisymmetric\/}, {\em stationary\/} and
{\em asymptotically flat\/}. We have further supposed spacetime to be
{\em circular\/}, thus the absence of meridional currents in the sources of
the gravitational field. In this case spacetime can be described by MSQI ({\em
Maximal Slicing\slsh Quasi Isotropic\/}) coordinates \cite{BGSM93} which have
a line element $ds^2\!=\!g_{\alpha\beta}\,dx^\alpha dx^\beta$ of the form:
\bgea ds^2=-N^2 dt^2+A^4 B^{-2}\,(dr^2\!+r^2 d\theta^2)\\
\nnum +A^4 B^2 r^2 \sin^2\!\theta\,(d\phi-N^\phi dt)^2.\edea
Spacetime is hence fully determined by the four metric potentials $N$,
$N^\phi$, $A$ and $B$. MSQI\slsh coordinates are global coordinates and lead
to elliptic operators which admit a consistent treatment of boundary conditions
and an efficient solution by spectral methods. The matter fields are chosen to
model a perfect fluid where first a polytropic, hence analytical, equation of
state was used. The assumption of a perfect fluid reduces the equations of
motion to an algebraic equation for the heat function $H$. From the Einstein
equations one derives four elliptic equations for the variables $\nu\!=\!\ln
N$, $N^\phi$ and
\bgeq G(r,\theta)=N(r,\theta)\,A^2(r,\theta)\,B(r,\theta),\edeq
\bgeq\zeta(r,\theta)=\nu(r,\theta)+2\alpha(r,\theta)-\beta(r,\theta)\edeq
whose final form reads
\bgeq\label{EQ:E1}\qquad\Delta_3\,\nu=\frac{A^4}{B^2}[4\pi(E\!+\!S^i{}_i)
+2(k_1^2\!+\!k_2^2)]-\partial\nu\,\partial(\nu\!+\!2\alpha\!+\!\beta),\edeq
\bgeq\label{EQ:E2}\qquad\tilde{\Delta}_3\,\tilde{N}^\phi=-16\pi\frac{N}{B^4}
\frac{J_\phi}{r\sin\theta}-r\sin\theta\,\partial N^{\phi} \partial(6\alpha\!
+\!3\beta\!-\!\nu),\edeq
\bgeq\label{EQ:E3}\Delta_2\,\tilde{G}=8\pi\frac{NA^6}{B}r\sin\theta\,
(S^r{}_r\!+\!S^\theta{}_\theta),\edeq
\bgeq\label{EQ:E4}\Delta_2\,\zeta = \frac{A^4}{B^2}[8\pi S^\phi{}_\phi+3
(k_1^2+k_2^2)]-(\partial\nu)^2\edeq
where we have introduced $\alpha\!=\!\ln A$, $\beta\!=\!\ln B$ and
\bgeq\quad\tilde{G}(r,\theta)=r\sin\theta\,G(r,\theta),\edeq
\bgeq\tilde{N}^\phi(r,\theta)=r\sin\theta\, N^{\phi}(r,\theta)\edeq
as well as the abridged notation 
\bgeq\partial\alpha\,\partial\beta=\partial_r\alpha\,\partial_r\beta
+\frac{1}{r^2}\,\partial_\theta\alpha\,\partial_\theta\beta.\edeq
Further employed quantities are the total energy density $E$, the stress tensor
$S^i{}_j$, the momentum density $J_i$ and $k_1$, $k_2$ which are related to the
extrinsic curvature tensor $K_{ij}$.
$\Delta_2$, $\Delta_3$ and $\tilde{\Delta}_3$ denote scalar Laplacians in two
and three dimensions and a vector Laplacian in three dimensions respectively.
\nspc
Existence and uniqueness of the solution of these elliptic equations are
ensured for physically relevant cases \cite{SY78-1,CA79,CA77}.\nspc
Note that (\ref{EQ:E3}) and (\ref{EQ:E4}) can be continued analytically to
yield genuine 2D\slsh Poisson equations in the entire $(r,\theta)$\slsh plane.
One infers immediately that $\zeta$ exhibits a logarithmic divergence for
$r\!\rightarrow\!\infty$ unless the total integral over the source of
(\ref{EQ:E4}) vanishes identically whereas we require $\zeta|_{\,r=\infty}
\!=\!0$. This 2D\slsh virial theorem of general relativity (GRV2)
\cite{BO73} is in some sense related to the classical Newtonian virial theorem
and furnishes a consistency condition for any solution of the Einstein
equations which is compatible with our basic assumptions. It has to be taken
into account during the calculation and provides a strong consistency check of
the numerical solution.
\subsection{Numerical method}\label{SUBSEC:NIX}
The mathematical problem involves (\ref{EQ:E1})\slsh (\ref{EQ:E4}) and an
algebraic first integral equation for the matter fields. Our numerical
solutions are exact in the sense that the governing equations are derived
from the full theory of general relativity without any analytical
approximation while the numerical code solves these equations in all space
extending the numerical integration to spatial infinity which allows to impose
the exact boundary conditions of asymptotical flatness on the gravitational
fields as well as the proper calculation of the source terms of
(\ref{EQ:E1})\slsh (\ref{EQ:E4}) which fill all space.\nspc
This is accomplished by the use of two grids, where the first one covers the
stellar interior using the radial variable $r$ in the interval $[0,R]$, while
the outer space is compactified thanks to the variable transform $u\!=\!r^{-1}$
and in this way mapping $[R,\infty]$ onto the finite interval $[R^{-1},0]$.
While in the $\theta-$variable a Fourier expansion according to
Sec.~\ref{SEC:CR} is used, the radial part is expanded in terms of Chebyshev
polynomials.
The inner zone is mapped onto half the definition interval $[0,1]$ which
allows to take into account the parity properties of regular functions with
respect to the origin according to Sec.~\ref{SUBSEC:AX}.
In the compactified zone the expansion has been done in the usual manner on
the whole definition interval $[-1,1]$ of Chebyshev polynomials.\nspc
The effective scheme which is based on a relaxation method works as follows.
We consider rigidly rotating neutron stars with a polytropic equation of state.
A particular configuration is hence determined by fixing the value $H_{\rm c}$
of the heat function at the centre of the star and its angular velocity
$\Omega$. 
We start from very crude initial conditions where all the metric quantities
are set to their flat space values ($\alpha$, $\beta$, $\nu$, $\zeta$ and
$N^{\phi}\!=\!0$; $G\!=\!1$) and the matter distribution is determined by a
first approximate guess.\nspc
While the GRV2 identity related to (\ref{EQ:E4}) holds for an exact solution
of the Einstein equations we have to enforce this consistency relation at each
iteration step in order to avoid a logarithmic divergence of the approximate
one.
This can be accomplished by modifying (\ref{EQ:E4}) according to
\bgeq\Delta_2\,\zeta=\sigma^{\rm f}_\zeta+\sigma^{\rm q}_\zeta\quad
\longrightarrow\quad\Delta_2\,\zeta=\sigma^{\rm f}_\zeta+\lambda
\sigma^{\rm q}_\zeta,\edeq
\bgeq\qquad\sigma^{\rm f}_\zeta=8\pi\frac{A^4}{B^2}\,S^\phi{}_\phi,\qquad
\sigma^{\rm q}_\zeta=\frac{A^4}{B^2}[3 (k_1^2\!+\!k_2^2)]-(\partial\nu)^2.\edeq
At each iteration step $\lambda$ is chosen in such a way that the total source
integral vanishes. The final solution has to satisfy (\ref{EQ:E4}) exactly
which is equivalent to $\lambda\!=\!1$. The deviation of $\lambda$ from unity
during the iteration measures the violation of self\slsh consistency of the
approximate solution and can be used to monitor convergence.\nspc
\begin{figure}[t]
\unitlength 1mm
\psfig{figure=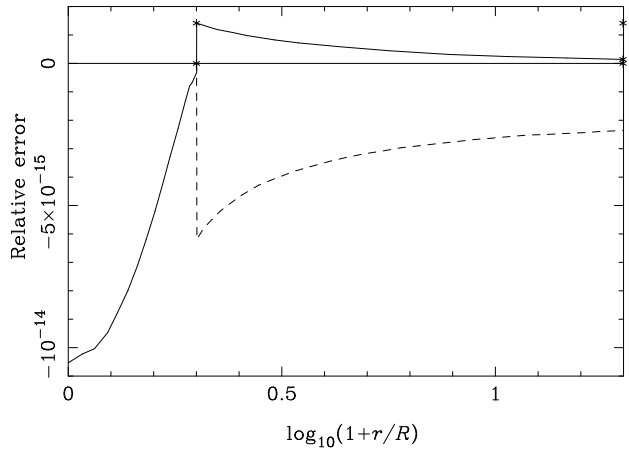,angle=270,width=95mm}
\caption[]{Comparison between numerical and analytical solution for the
Schwarzschild interior and exterior incompressible solution. The location of
the star surface is indicated by an asterisk. The plotted quantities are the
relative error in the pressure with respect to the central pressure inside the
star and the absolute error in $N$ (solid line) and $A^2$ (dashed line)
outside the star.}\label{FIG:SS}
\end{figure}
The sources of (\ref{EQ:E1})\slsh (\ref{EQ:E4}) exhibit some terms involving
simple operators like $r$, $\sin\theta$, $\partial_r$, etc. which are
accurately computed in coefficient space before evaluating the entire
expressions in configuration space. Expansion of the total sources in terms of
the angular eigenfunctions of the different Laplacians ($P^0_l(\cos\theta)$,
$P^1_l(\sin\theta)$ and $(\cos l\theta$, $\sin l\theta)$ for
$\Delta_3$, $\tilde{\Delta}_3$, and $\Delta_2$ respectively) leads to a system
of ODEs in the radial variable.
The unique global solution is obtained by appropriate linear combinations of
the corresponding particular and homogeneous solutions in order to match the
piecewise solutions at the grid interface and to satisfy the boundary
conditions at $r\!=\!\infty$. The new values of the gravitational field
variables are then used to update the matter distribution by means of the first
integral equation and the iteration can go on.
\subsection{Tests}
We have subjected our code to two different kinds of tests which ensure the
reliability of the numerical results.\nspc
{\em External\/} tests consist in the comparison with previous solutions,
either analytical or numerical ones.
Such a test of the code has been performed for an analy\-ti\-cally known
Schwarzschild type solution of a non\slsh rotating homogeneous sphere with the
corresponding numerical one. Relative errors committed on global quantities
such as total gravitational mass and circumferential radius are of the order of
$10^{-14}$.
This accuracy holds also for local quantities as shown in Fig.~\ref{FIG:SS}
for the pressure $p$ inside the star and the metric coefficients outside the
star --- none of the errors exceeding $10^{-14}$. A recent project of systematic
calibration and comparison of the numerical results of different groups
working in this field has yielded an agreement of characteristic quantities
of realistic neutron star models at a level of about $10^{-3}$.\nspc
{\em Internal\/} tests represent the second important class of tests and are
derived from some relations of global or local character which are related to
the Einstein equations but not automatically enforced during the calculation.
These tests are very powerful, since they do not only verify the proper working
of the numerical scheme for some simple --- usually degenerate --- test problem,
but apply to any calculation and supply an intrinsic estimate of the numerical
error involved. In the following neutron star matter was modeled by a pefect
fluid with a $\gamma\!=\!2$ polytropic EOS.
The angular velocity has been varied between $\Omega\!=\!0$ for the static case
and $\Omega\!=\!\Omega_{\rm K}$ for the maximum rotating case where for a
further increase of $\Omega$ mass shedding along the equator occurs.
Fig.~\ref{FIG:NS} shows the flattened shape of a neutron star rotating at
$\Omega_{\rm{K}}$.
\begin{figure}
\unitlength 1mm
\psfig{figure=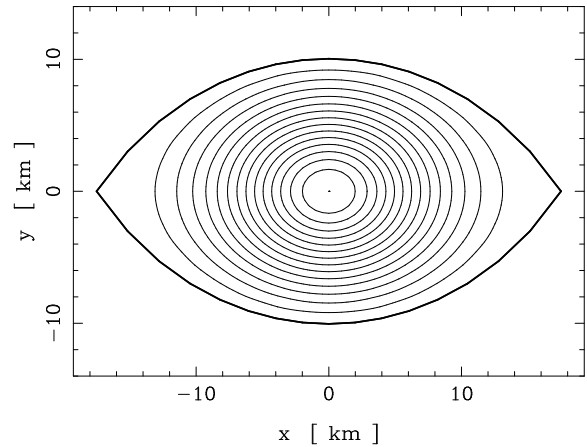,angle=270,width=95mm}
\caption[]{Level contour $E(r,\theta)$ in the case of a polytropic EOS with
$\gamma\!=\!2$ for $\Omega\!=\!\Omega_{\rm{K}}$.}\label{FIG:NS}
\end{figure}\hspace*{-0.5em}
A simple test makes use of the known analy\-ti\-cal expansion of $B$ according
to $B\!=\!1\!+\!r^2\sin^2\!\theta\,f$. It showed $B$ to coincide with the
analytical value 1 on the polar axis within $10^{-6}$ in the most unfavourable
case of maximal angular velocity. The principal test is provided by the GRV2
identity. An examination of the Schwarzschild type solution has revealed that
$|1\!-\!\lambda|$ is very closely related to the global errors derived from the
numerical solution for variable $N_r$. This observation, though obtained for
the static case, is supposed to hold in the rotating case as well.
$|1\!-\!\lambda|$ can thus be considered as an estimator of the global
numerical accuracy.
\begin{figure}[t]
\unitlength 1mm
\psfig{figure=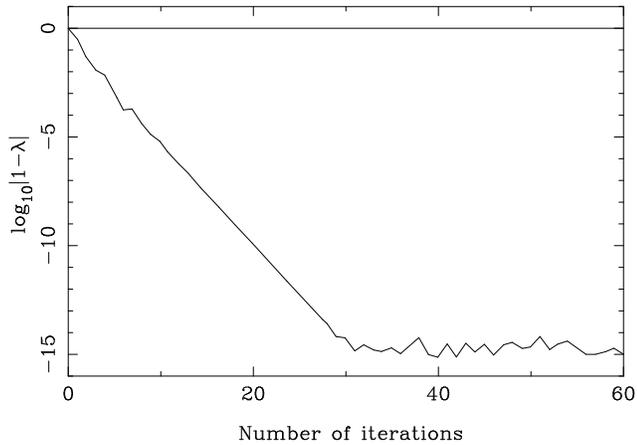,angle=270,width=95mm}
\caption[]{Convergence of the error indicator $|1\!-\!\lambda|$ during the
iteration process for $\Omega\!=\!0$.}\label{FIG:LBS}
\end{figure}
\begin{figure}[t]
\unitlength 1mm
\psfig{figure=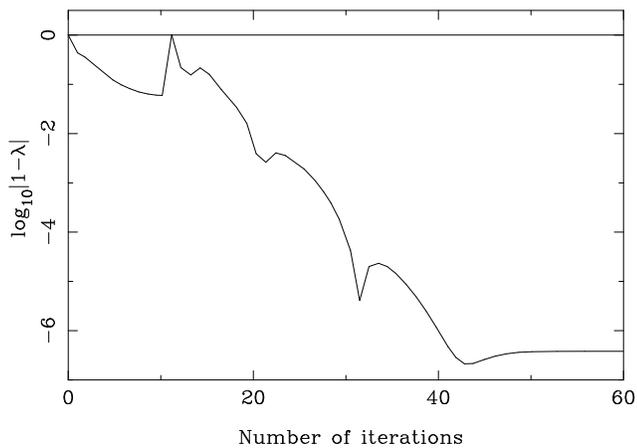,angle=270,width=95mm}
\caption[]{Convergence of the error indicator $|1\!-\!\lambda|$ during the
iteration process for $\Omega\!=\!\Omega_{\rm{K}}$. The spike at $N\!=\!10$ is
due to switching on the rotation.}\label{FIG:LBR}
\end{figure}\hspace*{-0.5em}
\begin{figure}[t]
\unitlength 1mm
\psfig{figure=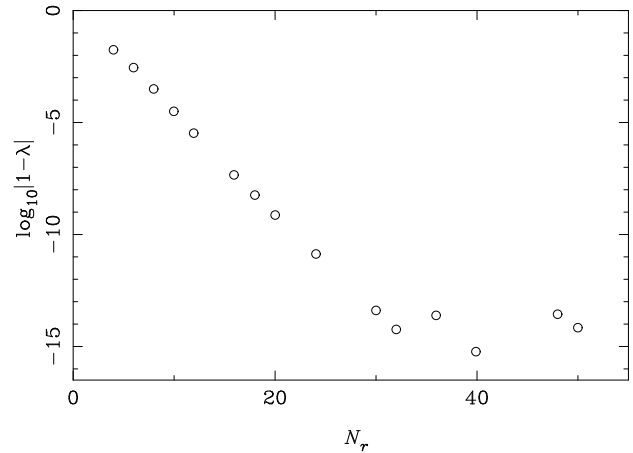,angle=270,width=95mm}
\caption[]{Internal error indicator as a function of $N_r$ in the compressible
spherically symmetric case (polytropic EOS).}
\label{FIG:ERE}
\end{figure}
\begin{figure}[t]
\unitlength 1mm
\psfig{figure=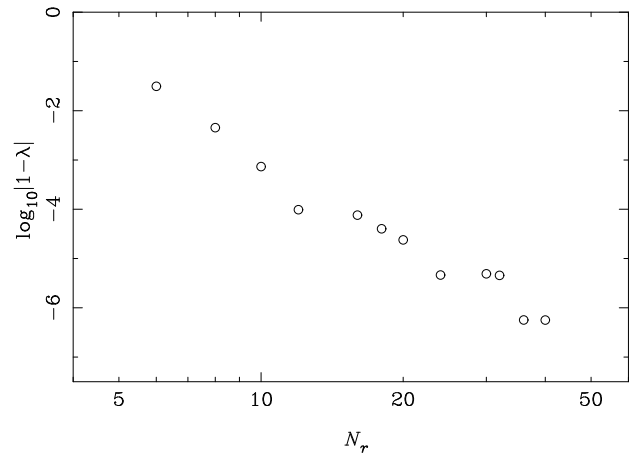,angle=270,width=95mm}
\caption[]{Same as Fig.~\ref{FIG:ERE} but for maximal angular velocity 
$\Omega_{\rm{K}}$. Pay attention to the $\log-\log$ scale; the spectral
properties are lost here and one observes a power law convergence.}
\label{FIG:ERX}
\end{figure}
Figs.~\ref{FIG:LBS}, \ref{FIG:LBR} show the convergence of $|1\!-\!\lambda|$
during the iteration for $\Omega\!=\!0$ and $\Omega\!=\!\Omega_{\rm{K}}$.
While in the first case the exponential decay of $|1\!-\!\lambda|$ continues
until roundoff errors of $10^{-14}$ mark a lower limit, the total error in
the rotating case is about $10^{-6}$.
This difference is due to the deviation of the flattened stellar shape from
the spherical numerical grid which leads to a discontinuity in the derivatives
across the stellar surface located {\em inside} the inner zone, where the
diverse quantities hence are no more analytical functions. The attainable
accuracy in dependence of the number of grid points is illustrated in
Fig.~\ref{FIG:ERE}.
The exponential decrease, usually called {\em evanescent\/} error and 
characteristic for spectral methods, is clearly visible. This property is lost
in the rotating case where we observe a power law decay of the committed error
$\propto\!N_r^{-4.5}$ as found from Fig.~\ref{FIG:ERX}. This inconvenience
will be overcome by the implementation of an adaptive ellipsoidal grid which
aligns the domain boundary to the star surface.
We finally conclude that we have computed neutron star models with an
analytic EOS achieving a precision of $10^{-14}$ in the static case to some
$10^{-6}$ in the maximum rotation case which has to be compared with previous
results based on finite difference methods of the order of $10^{-2}$
\cite{FIP86,KEH89a,KEH89b,CST92}.\nspc
It is further interesting to note that for typical values of $N_r\!=\!32$ and 
$N_\theta\!=\!16$ one iteration is performed in 480 ms on a VAX 4500. For an
average number of 50 iterations per model the whole calculation is finished in
about 24 s. The code efficiency has enabled us to carry out extensive studies
of neutron star samples under employment of numerous realistic EOS of neutron
star matter \cite{SBGH94}.
\subsection{Rotating neutron stars with magnetic field}
A further step toward a realistic description of rapidly rotating neutron stars
has been recently achieved by the fully self\slsh consistent inclusion of
magnetic fields into our models \cite{BBGN95}. These calculations represent the
first numerical solutions of the coupled 2D\slsh Einstein\slsh Maxwell
equations for rotating neutron stars which are hence fully relativistic,
taking into account any kind of interaction of the electromagnetic field with
the star and the gravitational field.\nspc
To complete the physical specification of our neutron star models as described
in Sec.~\ref{SUBSEC:NIX} we assume a perfect conductor behaviour of neutron
star matter --- the star interior is thus free of electric fields --- and add
the electromagnetic field variables $A_t$, $A_\phi$, the current variables
$j_t$ and $j_\phi$ and a structure function $f$ which determines the current
distribution inside the star. The additional free parameter which fixes a
unique neutron star configuration is the total electric charge $Q$.
A derived global quantity is the magnetic dipole moment $\cal{M}$ which
characterizes the magnetic properties of the neutron star.
The Maxwell equations lead to a set of coupled elliptic partial differential
equations which exhibits a similar structure as (\ref{EQ:E1})\slsh
(\ref{EQ:E4}). They involve a scalar Poisson equation for $A_t$ and a vector
Poisson equation for $A_\phi$ which read
\bgea\qquad\Delta_3\,A_t & = & -\mu_0\frac{A^4}{B^2}\,(g_{tt}j^t+g_{t\phi}
j^\phi)\\
\nnum & & -\frac{A^4 B^2}{N^2} N^\phi r^2 \sin^2\theta\times \partial A_t\,
\partial N^\phi\\
\nnum & & -\left(1+\frac{A^4 B^2}{N^2}\,(r\sin\theta N^\phi)^2\right)
\times\partial A_\phi\,\partial N^\phi\\
\nnum & & -(\partial A_t+2N^\phi\partial A_\phi)\,\partial(2\alpha\!+\!\beta
\!-\!\nu)\\
\nnum & &-2\,\frac{N^\phi}{r}\left(\partial_r A_\phi+\frac{1}{r\tan\theta}\,
\partial_\theta A_\phi\right),\edea
\bgea\qquad\tilde{\Delta}_3\,\tilde{A}^\phi & = & -\mu_0 A^8\,(j^\phi\!
-\!N^\phi j^t)\,r\sin\theta\\
\nnum & & +\frac{A^4 B^2}{N^2}\,r\sin\theta\,\partial N^\phi (\partial A_t
+N^\phi\partial A_\phi)\\
\nnum & & +\frac{1}{r\sin\theta}\,\partial A_\phi\,\partial(2\alpha\!+\!\beta\!
-\!\nu)\edea
where we define
\bgeq\tilde{A}^\phi=\frac{A_\phi}{r\sin\theta}.\edeq
\begin{figure}
\unitlength 1mm
\psfig{figure=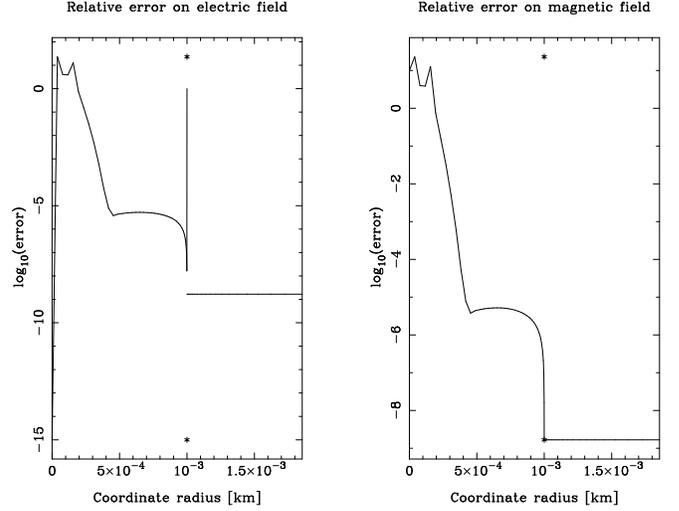,angle=270,width=95mm}
\caption[]{Comparison between the exact and the numerical solution in the case
of a rotating conducting sphere with a magnetic point dipole at its centre.
The asterisks indicate the sphere's surface.}
\label{FIG:DPT}
\end{figure}
In this extended formulation the determination of $A_t$ and $A_\phi$ precedes
the solution of the former set of equations.
$A_\phi$ is a smooth function in all space --- we recall that $\mu_r\!\simeq\!1$
for neutron star matter --- and can therefore be uniquely solved by imposing the
asymptotic boundary condition $A_\phi\!\rightarrow\!0$ for $r\!\rightarrow\!
\infty$.
The treatment of $A_t$ is slightly more complicated due to its non\slsh
smoothness across the stellar surface --- a behaviour caused by the surface
charges which are characteristic for ideal conductors. Since $A_t$ is linked
to $A_\phi$ by a linear relation in the star interior, the exterior solution
has to be matched to the latter one, respecting the condition $A_t\!=\!0$ at
infinity. 
We note that, once more, this is a task which is easily performed thanks to
the use of a spectral method. After solution of the whole system we can update
the electromagnetic contributions to the stress\slsh energy tensor and proceed
with the solution of the Einstein equations according to
Sec.~\ref{SUBSEC:NIX}.\nspc
Test calculations of the electromagnetic part have been performed for some
simple cases, where analytical solutions are known, among these one involving
a rotating magnetic dipole where an infinitesimal current loop at the origin
is surrounded by some rotating perfectly conducting sphere. This testbed
calculation mimics the configuration of a rotating neutron star with surface
charges.
Fig.~\ref{FIG:DPT} shows the relative error committed on the electric and the
magnetic field for a conducting sphere of $1\,\mbox{m}$ radius rotating at
$\Omega\!=\!3000$ s$^{-1}$ with a central current loop of $j_0\!=\!10^{11}$ 
Am$^{-2}$. Apart from the origin where the model current differs from the
theoretical $\delta$\slsh distribution the relative error is very small: about
$10^{-5}$ at half the radius and reaching its minimal value of $10^{-9}$
outside the sphere.
\begin{figure}
\unitlength 1mm
\psfig{figure=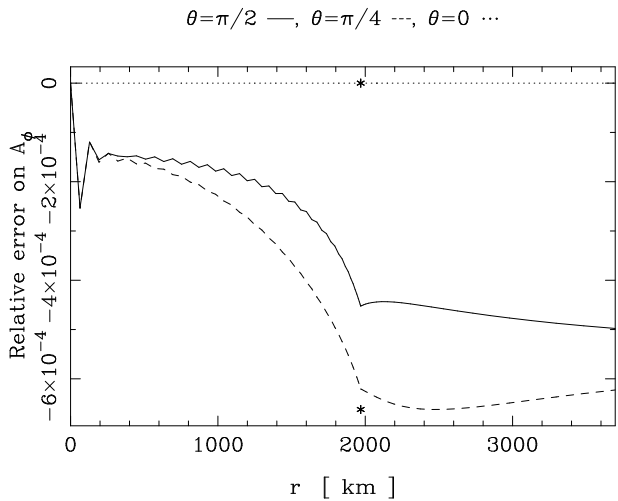,angle=270,width=95mm}
\caption[]{Comparison between Ferraro's analytical solution and the numerical
one in the case of a Newtonian incompressible fluid endowed with a magnetic
field corresponding to a constant current function $f(x)\!=\!f_0$. The plotted
quantity is the relative difference between the two values of $A_\phi$ as a
function of the radial coordinate $r$ for three values of $\theta$. Asterisks
indicate the star surface.}
\label{FIG:FPT}
\end{figure}
An analytical solution for a Newtonian incompressible fluid \cite{FE54} endowed
with a particular current distribution under the --- simplifying --- assumption
of spherical symmetry was adopted as a more sophisticated test. Also in this
case the agreement was quite good with a relative error of better than
$10^{-3}$, shown in Fig.~\ref{FIG:FPT}.
The deterioration with respect to the dipole problem is due to some simplifying
assumptions of the analytical model.\nspc
The accuracy of solutions of the complete Einstein\slsh Maxwell equations was
estimated as in Sec.~\ref{SUBSEC:NIX} by use of the virial identity GRV2
\cite{BG94} as well as by a more general three dimensional integral identity
valid for any stationary and asymptotically flat spacetime which we call GRV3
\cite{GB94}. It is the general relativistic generalization of the classical
Newtonian virial theorem. 
\begin{figure}[t]
\unitlength 1mm
\vspace*{-5mm}
\hspace*{-5mm}\mbox{\psfig{figure=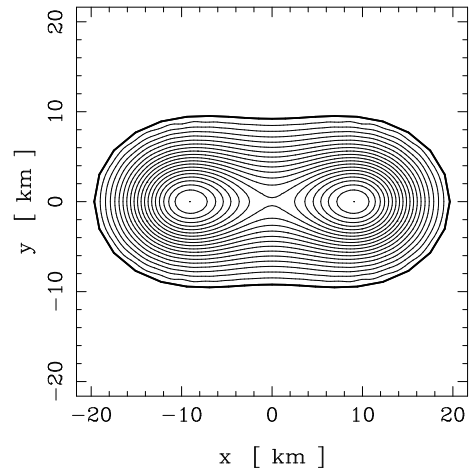,angle=270,width=100mm}}
\vspace*{-5mm}
\caption[]{Fluid proper density isocontours in the $(r,\theta)$\slsh plane for
the $M\!=\!4.06 M_\odot$ maximum mass static magnetized star built upon a
polytropic EOS for $\gamma\!=\!2$. The thick line indicates the star surface.}
\label{FIG:MMS}
\end{figure}
\begin{figure}[t]
\unitlength 1mm
\vspace*{-5mm}
\hspace*{-5mm}\mbox{\psfig{figure=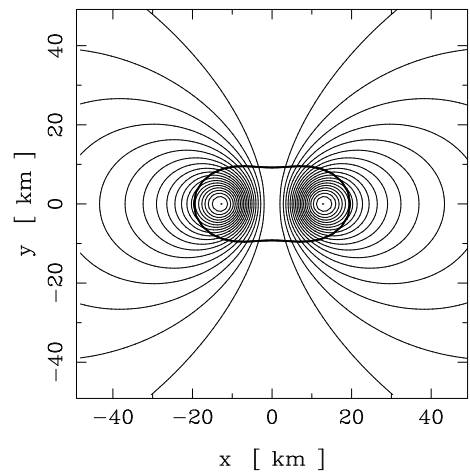,angle=270,width=100mm}}
\vspace*{-5mm}
\caption[]{Magnetic field lines in the $(r,\theta)$\slsh plane for the maximum
mass configuration corresponding to Fig.~\ref{FIG:MMS}. The thick line
indicates the star surface. The magnetic field amplitude amounts to
$B_{\mbox{c}}\!=\!9\!\times\!10^4$ GT at the star's centre.}
\label{FIG:MMF}
\end{figure}
The actual values of $|1\!-\!\lambda|$ showed about $10^{-5}$ for analytical
EOS and some $10^{-4}$ for the tabulated ones which agree with those of
calculations without magnetic field. Throughout all the calculations we chose
a grid resolution of 41 points in $r$ and 21 points in $\theta$.\nspc
In the following we studied configurations of {\em static\/} neutron stars
endowed with a magnetic field. The absence of kinematical effects allows an
unambiguous interpretation of the effects of the magnetic field.
In the static case the electric charge vanishes identically, leaving alone a
magnetic field. Since the stress\slsh energy tensor is {\em not\/} isotropic,
we observe a deformation of the star already in the static case.
Fig.~\ref{FIG:MMS} shows a maximum field configuration and Fig.~\ref{FIG:MMF}
the corresponding distribution of the magnetic field.
In the {\em rotating\/} case the magnetic field is accompanied by an additional
electric field. The both field distributions for a Pol2 $M\!=\!3.37 M_\odot$
model at $\Omega\!=\!3\!\times\! 10^3$ rad s$^{-1}$ and a constant current
function are given in Fig.~\ref{FIG:RFM}.
Note that the field lines of $A_t$ and $A_\phi$ coincide in the star interior
due to the perfect conductor assumption. The non\slsh smoothness of $A_t$
across the star boundary properly reflects the discontinuity of the electric
field due to the existing surface charges.
\begin{figure}[t]
\unitlength 1mm
\begin{center}
\vspace*{-18mm}
\hspace*{-4mm}\mbox{\psfig{figure=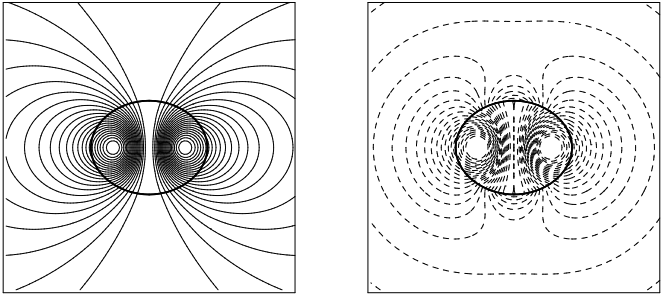,angle=270,width=97mm}}
\vspace*{-20mm}
\caption[]{a. Magnetic field lines in the $(r,\theta)$\slsh plane for the
configuration specified in the text. The thick line indicates the star
surface. b. Electric isopotential lines $A_t$=const.}
\label{FIG:RFM}
\end{center}
\end{figure}
\section{3D\slsh Gravitational neutron star collapse}
\subsection{Basic equations}\label{SEC:BHEM}
The investigation of the 3D\slsh gravitational collapse of rotating neutron
stars requires the solution of the general time\slsh dependent field equations
of general relativity. We stress that already in the Newtonian case a fully
three dimensional simulation of stellar collapse is a highly demanding and
non\slsh trivial problem.
Indeed up to this day there exists only one corresponding investigation which
aimed at the study of gravitational wave emission associated with type II
supernovae \cite{BM93}.
While in the presence of symmetries like stationarity or axisymmetry the field
equations are greatly simplified when evaluated explicitly for some appropriate
coordinates the situation is quite contrary for a three dimensional dynamical
problem. It is therefore favourable to solve the Einstein equations in their
original coordinate independent (covariant) form.
This choice allows us to adopt elliptic equations for the gauge variables $N$
and $N^i$ which have a direct geometrical signification.
The {\em maximal slicing\slsh minimal distortion gauge\/} \cite{SY78-1,SY78-2}
is highly singularity avoiding and neatly captures the propagation of
gravitational waves in the far field zone.
This is a very important feature, since the study of gravitational wave
emission is the principal goal of this investigation.\nspc
The governing equations then have the following form:
\bgeq\label{EQ:Q1}\partial_t h_{ij}\!+\!N_{i|j}\!+\!N_{j|i}\!+\!2 N K_{ij}
\!=\!0,\edeq\vspace*{-4.75ex}
\bgea\label{EQ:Q2}\partial_t K^i{}_j\!+\!N^l K^i{}_{j,l}\!-\!N^i{}_{,l}
K^l{}_j\!+\!N^l{}_{,j}K^i{}_l\\
\nnum\qquad\qquad +N^{|i}{}_{|j}\!-\!N[R^{i}{}_{j}\!+\!4\pi (S\!-\!E)\,
\delta^{i}{}_{j}\!-\!8\pi S^{i}{}_{j}]\!=\!0,\edea
\bgeq\label{EQ:Q3}R\!-\!K^i{}_j K^j{}_i\!-\!16\pi E\!=\!0,\edeq
\bgeq\label{EQ:Q4}K^j{}_{i|j}\!-\!8\pi J_{i}\!=\!0,\edeq
\bgeq\label{EQ:Q5}\partial_t E\!+\!N^{l}E_{,l}\!-\!N K^i{}_j S^j{}_i\!
+\!N^{-1}(N^2 J^i)_{|i}\!=\!0,\edeq
\bgeq\label{EQ:Q6}\qquad\partial_t J_i\!+\!N^l J_{i,l}\!+\!N^l{}_{,i}J_l\!
+\!(E\delta^{j}{}_{i}\!+\!S^j{}_i) N_{|j}\!+\!N S^j{}_{i|j}\!=\!0,\edeq
\bgeq\label{EQ:Q7}N^{|l}{}_{|l}\!-\!N[K^i{}_j K^j{}_i\!+\!4\pi(E\!+\!S)]\!=\!0,
\edeq
\bgeq\label{EQ:Q8}\qquad N^{i|j}{}_{|j}\!+\!\frac{1}{3}(N^j{}_{|j})^{|i}\!
+\!R^i{}_j N^j\!+\!2\,K^{ij}N_{|j}\!+\!16\pi N J^i\!=\!0.\edeq
This system of coupled partial differential equations is characterized by the
following properties:\nspc
The system includes a time first order hyperbolic system (\ref{EQ:Q1}) and
(\ref{EQ:Q2}) of the spatial metric tensor
$h_{ij}$ and its conjugate momentum variable $K_{ij}$ which reduces
to a wave equation for $h_{ij}$ in the far field zone.
The evolution of the matter fields $E$ and $J_i$ is governed by the parabolic
system (\ref{EQ:Q5}) and (\ref{EQ:Q6}). These equations constitute the
dynamical part of our problem.
$N$ and $N^i$ are subjected to the Poisson type equations (\ref{EQ:Q7}) and
(\ref{EQ:Q8}) where the matter fields act as source terms. Further involved
quantities are the stress tensor $S_{ij}$ and the Ricci tensor $R_{ij}$ where
$S\!=\!S^i{}_i$ and $R\!=\!R^i{}_i$.
The additional constraint equations (\ref{EQ:Q3}) and (\ref{EQ:Q4}) which
establish some consistency relations between gravitational and matter fields
are satisfied identically for the exact solution. They may be used to reduce
the number of the dynamical variables, an approach that would result in a
{\em constrained\/} evolution scheme. Indeed this procedure guarantees that
the numerical solution represents at any moment some solution of the Einstein
equations but not necessarily the correct one. We prefer an unconstrained
scheme where the constraint equations can serve to estimate the involved
numerical errors.
\subsection{Numerical method}\label{SUBSEC:NISF}
In the spirit of the analytical approach of Sec.~\ref{SEC:BHEM} we have opted
for a numerical scheme which is based on a one\slsh to\slsh one adaptation of
common tensor calculus, where elementary operations like contraction and
covariant derivation are performed by specific subroutines acting on entire
tensor quantities. We further introduce a flat background metric which enables
us to separate the contributions related to the curvature of space and to carry
out the numerical operations with respect to flat space spherical coordinates
and the corresponding metric tensor $f_{ij}$ \cite{RO63}.
For a given three\slsh metric $h_{ij}$ we then have the following
relations:\nspc
We first introduce a tensor $\Delta^i{}_{jk}$ defined as
\bgeq\Delta^i{}_{jk}=\frac{1}{2}\,h^{il}(h_{jl\|k}+h_{kl\|j}-h_{jk\|l})\edeq
where $_\|$ denotes covariant differentiation with respect to flat space
spherical coordinates. The covariant derivative $U^i{}_{|j}$ of a vector $U^i$
in curved three--space can then be reexpressed as
\bgeq U^i{}_{|j}=U^i{}_{\|j}+\Delta^i{}_{jk}U^k\edeq
where the generalization to tensors of higher order is obvious. We can further
rewrite the Ricci tensor $R_{ij}$ in terms of the $\Delta^i{}_{jk}$.
\bgeq R_{ij}=\Delta^l{}_{ij\|l}-\Delta^l{}_{jl\|i}+\Delta^l{}_{ij}
\Delta^m{}_{lm}-\Delta^l{}_{im}\Delta^m{}_{jl}.\edeq
The effective use of our scheme is clarified by inspection of the already
familiar scalar Poisson equation.
\bgeq N^{\mid i}{}_{\mid i}=S\quad\Longleftrightarrow\quad h^{ij}(N_{\|i\|j}
-\Delta^m{}_{ij} N_{\|m})=S.\edeq
Defining a new tensor field $\mbi{r}$ as $h^{ij}\!=\!f^{ij}\!+\!r^{ij}$ we can
isolate the $(f_{ij})$\slsh related covariant Laplacian and find the bimetric
equivalent of the covariant scalar Poisson equation
\bgeq N^{\|i}{}_{\|i}=S+(f^{ij}\!+\!r^{ij})\,\Delta^m{}_{ij} N_{\|m}-r^{ij}
N_{\|i\|j}.\label{EQ:DVDFN}\edeq
In the case of a conformally flat metric $h_{ij}\!=\!A^4 f_{ij}$ with $f_{ij}$
being the usual metric tensor of flat space spherical coordinates we obtain
$r^{ij}\!=\!(A^{-4}\!-\!1)\,f^{ij}$ and the following equation where $\Delta_f$
represents the usual flat space scalar Laplacian ($\alpha$ denotes $\ln A$).
\bgea \Delta_f N=S-2 A^{-4}(\partial_r\alpha\,\partial_r N+\frac{1}{r^2}\,
\partial_\theta\alpha\,\partial_\theta N\\
\nnum+\frac{1}{r^2\sin^2\theta}\,\partial_\phi\alpha\,\partial_\phi N)
-(A^{-4}\!-\!1)\,\Delta_f N.\edea
The linear contributions on the right can be eliminated, if we slightly change
the definition of $\mbi{r}$ to describe the deviation of the {\em conformal\/}
metric $\tilde{\mbi{h}}=\gamma^{-1}\mbi{h}$ from the flat space metric
$\mbi{f}$, where $\gamma^3=\mid\!\mbi{h}\!\mid\mid\!\mbi{f}\!\mid^{-1}$. So
with $\tilde{h}^{ij}\!=\!f^{ij}\!+\!r^{ij}$ we have $\mbi{r}\!=\!0$ and
\bgea \Delta_f N=A^4\,S-2\,(\partial_r\alpha\,\partial_r N+\frac{1}{r^2}\,
\partial_\theta\alpha\,\partial_\theta N\\
\nnum+\frac{1}{r^2\sin^2\theta}\,\partial_\phi\alpha\,\partial_\phi N).\edea
Though this result has been derived for a conformally flat metric we conclude
that for any three\slsh space the separation of the conformal factor is a first
improvement compared to the flat space approximation.\nspc
Following our reasoning in Sec.~\ref{SEC:CR} we restrict ourselves to the
exclusive use of {\em pseudophysical\/} components of tensor quantities related
to the standard local orthonormal frame of flat space spherical coordinates
in our numerical scheme.
The subroutines which calculate the covariant derivatives of tensors of order
zero to three appearing in our equations show the typical relative errors of
the order of $10^{-14}$ due to roundoff erors when using simple test functions.
For a successive derivation of a scalar function down to a tensor of order four
the relative error still does not exceed $10^{-12}$.
While the computation of the lapse equation can be reduced to the iterative
solution of a scalar Poisson\slsh like equation as demonstrated in
Sec.~\ref{SUBSEC:NIX}, there remain two other problems of higher demand. The
first one concerns the solution of the general shift vector equation involving
a linear vector operator which comprises a vector Laplacian and the gradient
of a divergence applied to the shift vector $N^i$. In Sec.~\ref{SUBSEC:VP} we
present a decomposition scheme based on the Clebsch\slsh Gordan theorem which
leads to an equivalent system of three scalar Poisson equations that can be
solved successively. The other one is related to the semi\slsh implicit time
integration of the evolution equations of the metric potentials. Here $R_{ij}$
includes a tensor Laplacian $\Delta h_{ij}$ which has to be treated implicitly.
An equivalent approach as in the vector case is in work. 
\subsection{Vector Poisson equation}\label{SUBSEC:VP}
The numerical inversion of a 3D\slsh vector Poisson equation is necessary to
solve the general shift vector equation which is an equation of the following
type
\bgeq \Delta\mbi{V}+\alpha\mbi{\nabla}(\mbi{\nabla}\!\cdot\!\mbi{V})=\mbi{S}
\label{EQ:SV}\edeq
where $\alpha$ is a constant. In order to facilitate the solution of
(\ref{EQ:SV}) we decompose $\mbi{V}$ and $\mbi{S}$ into its divergence\slsh
free and its irrotational part. We introduce vector fields $\tilde{\mbi{V}}$
and $\tilde{\mbi{S}}$ which we suppose to be divergence\slsh free as well as
two scalar potentials $\Psi$ and $\Phi$. We thus have a {\em unique}
decomposition
\bgea \label{EQ:DCV}\mbi{V}=\tilde{\mbi{V}}+\mbi{\nabla}\Psi\\
\label{EQ:DCS}\mbi{S}=\tilde{\mbi{S}}+\mbi{\nabla}\Phi\edea
after specifying the appropriate boundary conditions for $\Phi$ and $\Psi$.
In a first step we solve the Poisson equation for $\Phi$ which we obtain by
taking the divergence of (\ref{EQ:DCS}).
\bgeq\label{EQ:DP}\Delta\Phi=\mbi{\nabla}\!\cdot\!\mbi{S}.\edeq
Rewriting (\ref{EQ:SV}) in terms of the new variables we infer the equivalent
equation
\bgeq\label{EQ:SVD}\Delta\tilde{\mbi{V}}+\mbi{\nabla}((1\!+\!\alpha)\,\Delta
\Psi-\Phi)=\tilde{\mbi{S}}.\edeq
Taking the divergence of (\ref{EQ:SVD}) where we make use of the commutativity
of differential operators in flat space, we find a Poisson equation linking
$\Delta\Psi$ to the already known potential $\Phi$.
\bgeq\label{EQ:POT} (1\!+\!\alpha)\,\Delta\Delta\Psi=\Delta\Phi.\edeq
The solution of (\ref{EQ:POT}) is a priori determined up to an additional 
potential
$\Phi_{\rm{H}}$ with $\Delta\Phi_{\rm{H}}\!=\!0$. We fix $\Phi_{\rm{H}}$
to be zero according to the boundary conditions and obtain a Poisson equation
for $\Psi$.
\bgeq\label{EQ:POTT} (1\!+\!\alpha)\,\Delta\Psi=\Phi.\edeq
We turn now to (\ref{EQ:SVD}). Taking into account (\ref{EQ:POTT}) we derive
the final equation that governs the divergence\slsh free fraction of $\mbi{V}$.
\bgeq\label{EQ:VP}\Delta\tilde{\mbi{V}}=\tilde{\mbi{S}}.\edeq
We specify the components $V_i$ to be the physical components of the vector
$\mbi{V}$ related to the local orthonormal basis of spherical coordinates.
We further drop the tildes on the vector quantities. The explicit write\slsh
up of (\ref{EQ:VP}) then reads
\bgea\label{EQ:LPVR}\Delta_f V_r-\frac{2}{r^2}\,V_r
-\frac{2}{r^2}\left(\partial_\theta+\frac{1}{\tan\theta}\right)V_\theta\\
\nnum-\frac{2}{r^2\sin\theta}\,\partial_\phi V_\phi=S_r,\edea
\bgea\label{EQ:LPVT}\Delta_f V_\theta-\frac{1}{r^2\sin^2\theta}\,V_\theta
+\frac{2}{r^2}\,\partial_\theta V_r\\
\nnum -\frac{2}{r^2\sin\theta\tan\theta}\,\partial_\phi V_\phi=S_\theta,
\edea
\bgea\label{EQ:LPVF}\Delta_f V_\phi-\frac{1}{r^2\sin^2\!\theta}\,V_\phi
+\frac{2}{r^2\sin\theta}\,\partial_\phi V_r\\
\nnum+\frac{2}{r^2\sin\theta\tan\theta}\,\partial_\phi V_\theta=S_\phi
\edea
where $\Delta_f$ denotes the ordinary scalar Laplacian
\bgeq\label{EQ:LP}\qquad\Delta_f=\partial^2_r+\frac{2}{r}\,\partial_r
+\frac{1}{r^2\tan\theta}\,\partial_\theta+\frac{1}{r^2}\,\partial^2_\theta
+\frac{1}{r^2\sin^2\!\theta}\,\partial^2_\phi.\edeq
We further recall the representation of the covariant divergence 
$\mbi{\nabla}\!\cdot\!\mbi{S}$ in spherical coordinates
\bgea\label{EQ:DVS}\mbi{\nabla}\!\cdot\!\mbi{S}=\partial_r S_r+\frac{2}{r}
\,S_r+\frac{1}{r}\,\partial_\theta S_\theta+\frac{1}{r\tan\theta}\,S_\theta\\
\nnum+\frac{1}{r\sin\theta}\,\partial_\phi S_\phi.
\edea
We adopt the following notation where we introduce two auxiliary scalar
potentials $U$ and $W$ in order to obtain a set of decoupled equations which
is equivalent to (\ref{EQ:LPVR})\slsh (\ref{EQ:LPVF}).
\bgea V_\theta=\frac{1}{r}\,\partial_\theta U-\frac{1}{r\sin\theta}\,
\partial_\phi W,\\
V_\phi=\frac{1}{r}\,\partial_\theta W+\frac{1}{r\sin\theta}\,\partial_\phi U.
\label{EQ:AP}\edea
From $\mbi{\nabla}\!\cdot\!\mbi{V}\!=\!0$ we get
\bgea\label{EQ:DVV}\partial_r V_r+\frac{2}{r}\,V_r+\frac{1}{r^2}\,
\partial^2_\theta U+\frac{1}{r^2\tan\theta}\,\partial_\theta U\\
\nnum+\frac{1}{r^2\sin^2\!\theta}\,\partial^2_\phi U=0.
\edea
Combining (\ref{EQ:DVV}) with (\ref{EQ:LPVR}) leads us to a scalar Poisson
equation for $\tilde{V}$ where we have defined $\tilde{V}\!=\!rV_r$.
\bgeq\label{EQ:TV}\Delta_f\tilde{V}=r S_r.\edeq
Proceeding in the same manner with (\ref{EQ:LPVT}) and (\ref{EQ:LPVF}) we
derive the following linear combinations of the resulting equations with
(\ref{EQ:DVV}).
\bgea\label{EQ:IM1}\sin\theta\,\partial_\theta(\partial^2_r U\!-\!\partial_r
V_r)-\partial_\phi\left(\Delta W-\frac{2}{r}\,\partial_r W\right)\\
\nnum =r\sin\theta\,S_\theta,\\
\label{EQ:IM2}\frac{1}{\sin\theta}\,\partial_\phi(\partial^2_r U\!
-\!\partial_r V_r)+\partial_\theta\left(\Delta W-\frac{2}{r}\,\partial_r
W\right)\\\nnum=r\,S_\phi.\edea
Derivation of (\ref{EQ:IM1}) with respect to $\theta$ and of (\ref{EQ:IM2})
with respect to $\phi$ and adding the both equations allows us to recover a
scalar Poisson equation for $(\partial_r U\!-\!V_r)$ which only involves the
angular variables $(\theta,\phi)$ where we have integrated over $r$ and set
the implied integration constant to zero to assure a vanishing behaviour of
the source terms at $r\!=\!\infty$.
\bgeq\Delta_{\theta\phi} (\partial_r U\!-\!V_r)=-r^2 S_r.\edeq
Here $\Delta_{\theta\phi}$ denotes the angular fraction of $\Delta_f$
multiplied by $r^2$.\nspc
At this point we can determine $U$ by means of $V_r$ which has already been
fixed by (\ref{EQ:TV}).
In order to calculate the lacking potential $W$ we take up (\ref{EQ:IM2}).
An ordinary integration over $\theta$ where we require vanishing behaviour of
the source terms at infinity as for $(\partial_r U\!-\!V_r)$ results in a
scalar Poisson equation for $\tilde{W}$ defined by $W\!=\!r\tilde{W}$,
\bgeq r\Delta \tilde{W}=\int_0^\theta\left[r S_\phi-\frac{1}{\sin\theta}\,
\partial_\phi (\partial^2_r U\!-\!\partial_r V_r)\right]d\theta.\edeq
This scheme hence allows to compute the required potentials successively for
a given source distribution starting from the ordinary Poisson equation for
$\tilde{V}$, solving then the equation involving $(\partial_r U\!-\!V_r)$ and
thus fixing $U$ while in a last step $\tilde{W}$ can be determined as a
quantity depending on $V_r$ and $U$. Note that $S_\theta$ does not appear in
the source terms of the final equations. This is due to the constraint equation
$\mbi{\nabla}\!\cdot\!\mbi{S}\!=\!0$ removing one degree of freedom.\nspc
\begin{figure}[t]
\unitlength 1mm
\psfig{figure=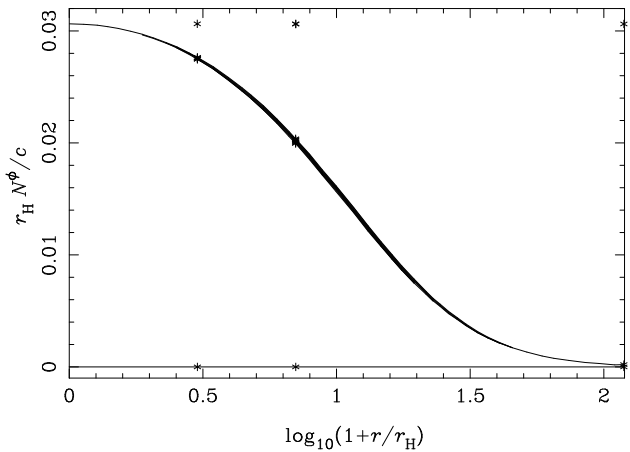,angle=270,width=95mm}
\caption[]{$N^\phi$ for a rapidly rotating Kerr black hole at $a/M\!=\!0.99$
where $r_{\rm H}$ denotes the radius of the horizon. The different curves
correspond to various values of $\theta$ ranging from $0^\circ$ to $90^\circ$.
The asterisks indicate the subdomain boundaries.}
\label{FIG:KNFF}
\end{figure}
\begin{figure}[t]
\unitlength 1mm
\psfig{figure=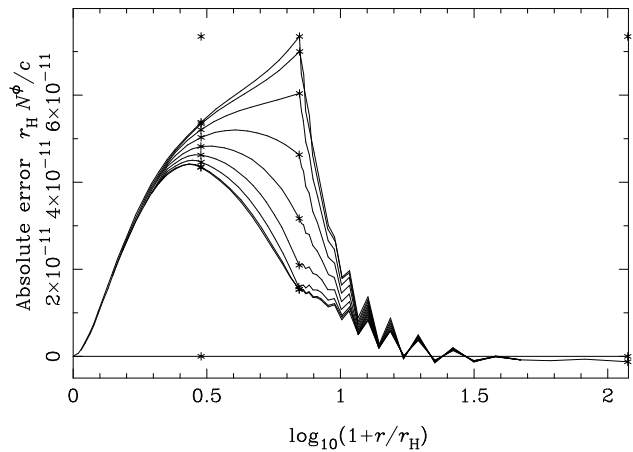,angle=270,width=95mm}
\caption[]{Absolute error in $N^\phi$ corresponding to Fig.~\ref{FIG:KNFF}.}
\label{FIG:KNFE}
\end{figure}
\subsection{Tests}
The routines we have built based on this computation scheme and on our spectral
method library enable us to solve a vector Poisson equation in a multidomain
configuration including an exterior compactified zone --- if desired --- which
covers all space and thus allows to impose proper boundary conditions for
asymptotically flat space where for the time being we ristrict ourselves to
the supersymmetric case.
Numerical tests have been performed on simple test functions and on problems
where the analytical solution was known in advance. For test functions we found
once more the numerical error to be governed by the roundoff limit of the
employed machine of the order of $10^{-14}$.
As an advanced test problem we solved the shift vector equation in the Kerr
metric of a rotating black hole. The presented configuration corresponds to a
rapidly rotating Kerr hole --- $a/M\simeq 0.99$ --- close to maximum angular
velocity where the relativistic effects involving the shift vector component
$N^\phi$ are strongly pronounced. The entire space outside the black hole is
covered by three zones where the outer compactified one extends to spatial
infinity. The grid resolution is chosen to be $N_r\!=\!33$ in each zone and
$N_\theta\!=\!9$.
Note the quite small number of nodes in $\theta$. Thanks to taking into account
the even symmetry of the problem with respect to reflection at the equatorial
plane according to Sec.~\ref{SUBSEC:SS} this corresponds to an effective value
of 17 in the whole interval $[0,\pi]$.
Fig.~\ref{FIG:KNFF} shows the course of $N^\phi$ in the vicinity of the black
hole while Fig.~\ref{FIG:KNFE} illustrates the absolute error committed on the
shift vector component $N^\phi$. The numerical error nowhere exceeds
$10^{-10}$. For a higher value of $N_r$ it even goes down to about $10^{-13}$.
As expected the numerical errors are most elevated at the boundaries of the
different subdomains where the piecewise solutions of adjacent shells are
matched. The GRV2 identity which had already proved its usefulness in the
study of axisymmetric stationary neutron stars was applied to estimate the
total error of the numerically computed Kerr spacetime. The error estimator 
$|1\!-\!\lambda|$ turned out to be closely related to the numerical errors
derived above from comparison with the analytical solution and confirmed in
this highly relativistic problem to be a sensitive indicator of the global
numerical accuracy.
\section{Conclusion}
We have presented the application of spectral methods to several problems of
numerical relativity. In each case they proved to be a highly valuable tool
which lead to results typically several orders of magnitude more accurate than
corresponding codes based on finite difference schemes. Especially in
sphericallike coordinates the advantages of a spectral method which allows a
rigorous treatment of the associated regularity conditions, while improving
the efficiency of the code at the same time, are remarkable. Particularly
important properties for our problems are the negligible numerical viscosity
in temporal evolution schemes which enabled us to capture subtle details in
the time\slsh dependence of evolved variables as observed for equilibrium
configurations of neutron stars in Sec.~\ref{SUBSEC:NSR}, as well as the very
natural treatment of boundary conditions and the efficient solution of
elliptic equations which is a frequently encountered task in our
investigations.
Our so far very positive experiences with spectral methods give us confidence
to dispose of the appropriate numerical tool to tackle the exciting problem of
black hole formation by 3D\slsh gravitational collapse of neutron stars.
\\[7.5ex]
{\em Acknowledgements.} J. Frieben gratefully acknowledges financial support
by the {\sc Gottlieb Daimler\slsh und Karl Benz\slsh Stiftung}.
\end{document}